\documentclass{article}

\usepackage{PRIMEarxiv}

\usepackage[utf8]{inputenc} 
\usepackage[T1]{fontenc}    
\usepackage{hyperref}       
\usepackage{url}            
\usepackage{booktabs}       
\usepackage{amsfonts}       
\usepackage{nicefrac}       
\usepackage{microtype}      
\usepackage{lipsum}
\usepackage{fancyhdr}       
\usepackage{graphicx}
\usepackage{amsmath}
\usepackage{amssymb}
\usepackage{booktabs}
\usepackage{float}

\title{Predicting Regional Road Transport Emissions From Satellite Imagery}

\author{
  Adam Horsler\\
  Imperiall College London\\
  London\\
  \texttt{adam.horsler19@alumni.imperial.ac.uk}
  \\
   \And
    Jake Baker \\
  Imperial College London\\
  London\\
  \texttt{jake.baker18@alumni.imperial.ac.uk} \\
  \And
    Pedro M. Baiz. V.\\
  Imperial College London\\
  London\\
  \texttt{p.m.baiz@imperial.ac.uk}\\
}

\begin{document}
\maketitle

\begin{abstract}
This paper presents a novel two-part pipeline for monitoring progress towards the UN Sustainable Development Goals (SDG's) related to Climate Action and Sustainable Cities and Communities. The pipeline consists of two main parts: the first part takes a raw satellite image of a motorway section and produces traffic count predictions for count sites within the image; the second part takes these predicted traffic counts and other variables to produce estimates of Local Authority (LA) motorway Average Annual Daily Traffic (AADT) and Greenhouse Gas (GHG) emissions on a per vehicle type basis. We also provide flexibility to the pipeline by implementing a novel method for estimating emissions when data on AADT per vehicle type or/and live vehicle speeds are not available. Finally, we extend the pipeline to also estimate LA A-Roads and minor roads AADT and GHG emissions. We treat the 2017 year as training and 2018 as the test year. Results show that it is possible to predict AADT and GHG emissions from satellite imagery, with motorway test year $R^2$ values of 0.92 and 0.78 respectively, and for A-roads' $R^2$ values of 0.94 and 0.98. This end-to-end two-part pipeline builds upon and combines previous research in road transportation traffic flows, speed estimation from satellite imagery, and emissions estimation, providing new contributions and insights into these areas.
\end{abstract}

\keywords{Machine Learning for Satellite Imagery \and Road Transport \and Greenhouse Gas Emissions}

\section{Introduction}
Satellite imagery has emerged as a promising data source for addressing challenges in quantifying GHG emissions progress, such as regional regulatory differences and limited data availability. It provides an unbiased and non-invasive source of information that can be standardized and compared across regions. This project leverages satellite imagery and public and free datasets with Deep Learning (DL) methods to estimate road vehicle greenhouse gas (GHG) emissions by predicting traffic flows from a single satellite image. Transportation is a significant contributor to GHG emissions, accounting for 14\% of all direct and indirect emissions. Road transportation is the largest component of this sector, responsible for 73\% of total emissions \cite{lamb2021review}, hence the focus on road transport within this study. 
 
This work develops a novel pipeline for transforming a single satellite image into road transport Average Annual Daily Traffic (AADT) and GHG emissions. To that end, the following intermediate steps are required:
satellite image pre-processing and metadata extraction, road vehicle object detection and traffic counts prediction. 
 
While the scope of this project is limited to the UK, additional work provides a flexible method that can be adapted to different regions, specifically those with limited data availability compared to the UK. For example, we incorporate the optional method of live speed estimation for estimating traffic counts when ground truth data is unavailable. This is in relation to the aforementioned SGD's, which require a global effort for progress to be most impactful. 
\begin{figure*}
    \centering
    \includegraphics[width=1\textwidth]{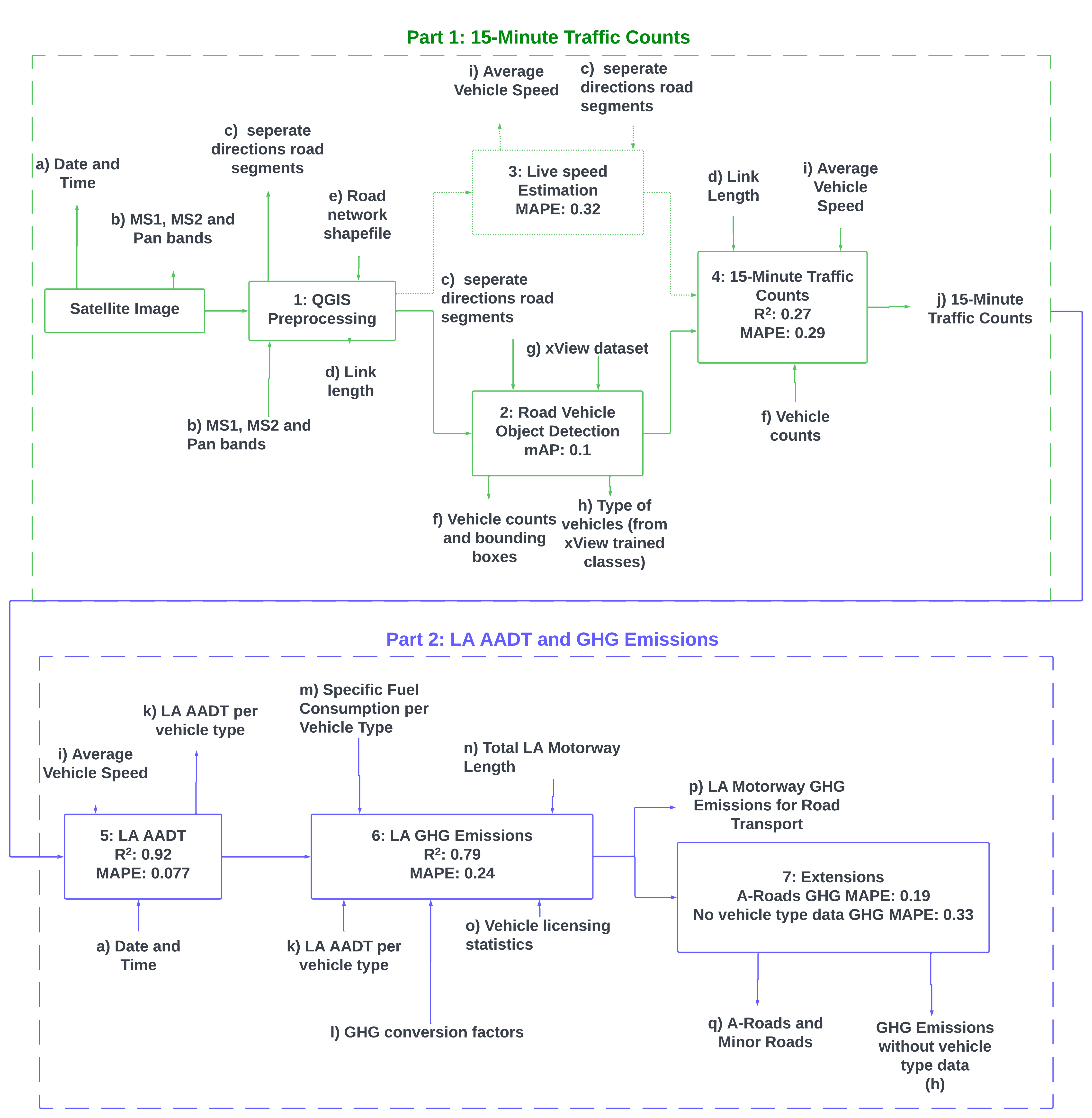}
    \caption{Full Pipeline Diagram}
    \label{fig:full_pipeline_diagram}
\end{figure*}
 
The full pipeline diagram is shown in Figure \ref{fig:full_pipeline_diagram}. The diagram is split into two parts: 15-minute traffic counts (green) and LA AADT and GHG emissions (blue), and within each part are stages that are numbered and within rectangular blocks. Live speed estimation is dotted to reflect the fact that it is a flexible step in the pipeline. The variables are lettered, e.g. a), b) etc. for readers to better follow their usage throughout the pipeline diagram and arrows going into a rectangle represent inputs, while arrows going outwards represent outputs. Finally, we include our key metric results from relevant stages in the diagram. Part 1 results are found in Section \ref{Results: 15-Minute Traffic Counts} and Part 2 results are found in Sections \ref{Results: Road Transport LA AADT} and \ref{Results: Road Transport LA GHG Emissions}.

\section{Related Work}

We provide an overview of methods that use satellite imagery to develop GHG emissions prediction of road transport. Due to the nature of this work, there are currently no previous works that use the same methods throughout the full pipeline, therefore we split this section into sub-sections that correspond to the major parts of the pipeline. 

\subsection{Part 1: 15-Minute Traffic Counts} \label{Modelling Methodologies: Satellite Imagery Pre-Processing}

Stage 1, satellite imagery pre-processing, is important for accurate vehicle detection in satellite imagery. It involves essential pre-processing steps aimed at isolating the target road segment from the raw image while eliminating extraneous vehicles like parked cars. Techniques such as pan-sharpening, which merges high-resolution panchromatic and lower-resolution multi-spectral images are used to improve results \cite{li2017assessment}. Additionally, more advanced approaches such as a Line Segment Detector (LSD) in conjunction with road direction information, as demonstrated by Ganji et al. \cite{ganji2022traffic} have been researched. Alternatively, deep learning techniques like semantic segmentation are becoming more used, with models like Zhu et al.'s Global Context-aware and Batch independent Network (GCB-Net) \cite{zhu2021global} and the widely accessible DeepLabv3 \cite{chen2017rethinking} that automate road segment extraction by masking background pixels.
 
In Stage 2, satellite-based vehicle detection using DL, Convolutional Neural Network (CNN) models like YOLO, SSD, RetinaNet, and Faster R-CNN are commonly employed due to their superior performance. However, this task faces challenges related to satellite imagery, notably small road vehicle objects in comparison to image size, complex backgrounds and variations in weather and occlusion. Maity et al. \cite{9418274} initially favored YOLOv5 for this purpose, but YOLOv8 has since emerged as an improvement. Gong et al. \cite{gong2022swin} introduced SPH-YOLOv5 with Swin Transformer Prediction Heads, which demonstrated slight performance enhancements on popular datasets like DOTA \cite{xia2018dota}. Addressing the issue of small object detection, SAHI \cite{akyon2022sahi} proposes Slicing Aided Hyper Inference and Fine-tuning, a method specifically effective for satellite imagery due to its ability to handle small object resolutions.
 
Stage 3, live speed estimation, is optional for when true speed data at the time of satellite image acquisition is unavailable. Vehicle speeds at the time of satellite image capture play a crucial role in estimating traffic counts, as well as subsequent metrics like AADT and GHG emissions. Higher average vehicle speeds during image acquisition indicate increased road traffic flow, while lower speeds imply fewer vehicles passing through. Salehi et al. \cite{6145661} proposes an innovative approach by introducing a method for estimating road vehicle speeds from WorldView-2/3 satellite images, leveraging the 0.26-second time gap between MS1 and MS2 band images. This technique, based on Principal Component Analysis (PCA), requires no additional data beyond what is already available from a WorldView-2/3 satellite image.
 
For stage 4, predicting traffic counts, numerous approaches have emerged to convert vehicle detection in satellite/aerial imagery into hourly and daily traffic counts, with a common successful feature being the segmentation of roads into individual directions to account for varying traffic patterns. Ganji et al. \cite{ganji2022traffic} employ a simple artificial neural network (ANN) with one hidden layer to predict daily traffic counts based on inputs like vehicle detection count, road characteristics, and aerial image timestamps. Road characteristics encompass factors such as road width, type, lane numbers, and speed limits, making it a relatively intricate method due to substantial pre-processing requirements for each input. Kaack et al. \cite{10.1145/3314344.3332480} propose a straightforward formula for converting vehicle detections and road speeds into daily traffic counts by considering the number of detected vehicles and their constant speed within a defined interval:
\begin{equation}
\text{Daily Traffic Count} = c_I \frac{24}{t_I}
\end{equation} 
Blattner et al. \cite{blattner2021commercial} adopts a regression tree-based model, CatBoost \cite{dorogush2018catboost}, for predicting hourly traffic counts, including inputs like weekday and cloud coverage over freeway areas. Additional section-specific features, such as lane count and proximity to large cities, were tested but did not enhance model performance. Importantly, many of these methods rely on true speed data or estimates. 

\subsection{Part 2: LA AADT and GHG Emisisons}

Stage 5 converts traffic counts into AADT estimations. Ganji et al. \cite{ganji2022traffic} applies an approach reminiscent of their prior Pattern Recognition Traffic Counts (PRTC) method \cite{ganji2020methodology}, utilizing nearby traffic count site data to estimate Annual Average Daily Traffic (AADT) for bidirectional vehicle count detections. Kaack et al. \cite{10.1145/3314344.3332480} converts daily traffic counts to AADT, factoring in traffic density variations across hours, days, and months, with AADT approximated as Daily Traffic Counts multiplied by 24 hours divided by time taken for vehicle travel and the inverse of the density factor. They also normalize count values by dividing hourly counts by their annual mean. The formula is as follows:
\begin{equation}
\text{AADTT} = c_I \frac{24h}{t_I} f^{-1}_{h,d,m} = \text{Daily Traffic Counts} \cdot f^{-1}_{h,d,m}
\end{equation}
Conversely, the UK Department for Transport (DfT) employs a method tailored to available vehicle activity data, utilizing expansion factors to convert 12-hour manual counts into AADT.
 
Stage 6 estimates GHG emissions from AADT and other correlated variables. Ganji et al.'s 2020 paper \cite{ganji2020methodology} extended their AADT prediction method to simultaneously forecast GHG emissions on individual roads over multiple years across an extensive road network. They calculated segment emissions by multiplying segment-level vehicle kilometers traveled (VKT) with corresponding emission factors (in g/veh km). Estimating road speed data at various times of day and week was crucial for determining these emission factors, which they achieved using an artificial neural network (ANN) with inputs correlated to road speed. To analyze the correlations between GHG emissions, AADT, and speed, the results were normalized by road length.

\section{Data \& Methods}

This section will outline the datasets and how we use them in the models for each part of the pipeline, the end goal being to produce LA-level road transport AADT and GHG emissions from a single satellite image. We also introduce several flexible approaches to vehicle speed and type data for when these datasets are not available. Finally, we describe the method behind extending the pipeline to also estimate LA AADT and GHG emissions for A-roads and minor roads. 

\subsection{Road Vehicle Object Detection}
This section represents stages 1 and 2 of the pipeline: QGIS Pre-Processing and Road Vehicle Object Detection, from Figure \ref{fig:full_pipeline_diagram}.
 
The input to the pipeline is a satellite image covering the chosen count sites. If vehicle speed estimation is required, the 8-band WorldView-2/3 satellite image is required. The image acquisition date and time are extracted for use in AADT prediction. 
 
QGIS pre-processing is used to convert the satellite image into a Pan-sharpened image with separate directions and a masked background. The rest of the pipeline uses only one direction, as it is repeated for the other direction. Thus, each image input will produce 2 masked output images for each direction. The following steps describe the data pre-processing:
\begin{enumerate}
    \item Load raw image and import road network
    \item Project the datasets onto the same co-ordinate system
    \item Create a Pan-Sharpened image using the RGB bands of MS1 and grayscale Pan images. 
    \item Create a buffer around the road network that includes only the bi-directional road width.
    \item Intersect the buffer with the satellite images
    \item Save the resulting intersected images
\end{enumerate}

The processed Pan-sharpened image is then used for vehicle object detection. For road vehicle object detection, we use the YOLOv5 medium-sized model trained on the road vehicle classes of the xView dataset \cite{lam2018xview}. Training uses the Ultralytics \cite{Jocher_YOLOv5_by_Ultralytics_2020} and small object detection \cite{akyon2022sahi} frameworks. 
 
A significant change made to the YOLOv5 Ultralytics default configuration is the incorporation of the focal loss function from RetinaNet \cite{lin2017focal}. This is due to the class imbalances in the xView dataset and complex background with small objects. The inference output for each detected vehicle and its usage in the pipeline is as follows:
\begin{itemize}
    \item bounding box values ($x\_min$, $x\_max$, $y\_min$, $y\_max$) for calculating vehicle length for compatibility with the UK traffic counts dataset \cite{HighwaysEnglandTrafficFlowData}
    \item xView vehicle classification for use when AADT vehicle type data isn't available. The mapping table and implementation are described in Section \ref{Methods: Without Vehicle Type Data}
\end{itemize}

\subsection{Motorway Traffic Counts} \label{chapter: Implementation of Motorway Traffic Counts}

This section represents stage 4 of the pipeline, from Figure \ref{fig:full_pipeline_diagram}. An estimation of 15-minute traffic counts is made using the output of the object detection model, the road length segment of the processed satellite image, and speed data. The speed data can be either true speeds, as readily available in the UK \cite{HighwaysEnglandTrafficFlowData} or estimated with the method described in Section \ref{Methods: Live Speed Estimation} 
 
We extend the method by Kaack et al. \cite{10.1145/3314344.3332480} by categorising traffic counts by vehicle length. This is to be compatible and make full use of the UK ground truth dataset provided by Highways England \cite{HighwaysEnglandTrafficFlowData}.
\begin{equation}
    N_{i15} = 15 \frac{vN_i}{l}
\end{equation}


\subsection{Motorway LA AADT} \label{Implementation: Local Authority AADT Prediction}

In this section, we discuss the method of converting the 15-minute traffic count estimates into motorway LA AADT estimations. This represents stage 5 of the pipeline. 

\subsubsection{ANN AADT Pre-Processing and Training}

The Highways England traffic count dataset \cite{HighwaysEnglandTrafficFlowData} is first pre-processed to be suitable for input into an ANN. Each row in the dataset for ANN AADT Predictions is treated as a different data-point, thus assuming independent and identically distributed data. A min-max standardisation step is applied on the vehicle count features for each site. Min-max is chosen as this ensures the minimum value for vehicle counts will be 0 and standardisation has been shown to improve performance \cite{ganji2022traffic}. This standardisation is not done on the other features as they are time-based. The ANN is a multi-input and multi-output system, where the inputs are summarised in Table \ref{table:ANN_AADT_Features}

\begin{table}[H]
  \centering
  \begin{tabular}{@{} lcr @{}}
\toprule
Feature & Description\\
\midrule
Hour of Day & The hour of the day\\ 
\hline
Day of Week & The day of the week \\ 
\hline
Month of Year & The month of the year\\ 
\hline
Live Speed & Average vehicle speed\\ 
\hline
Small vehicle count & Number of vehicles \\ between 0-520cm\\
\hline
Medium vehicle count & Number of vehicles \\ between 521-660cm\\
\hline
Large vehicle count & Number of vehicles \\ between 661-1160cm\\
\hline
Very Large vehicle count & Number of vehicles \\ over 1160cm\\
\hline
Total Vehicles & Total number of vehicles\\
\bottomrule
\end{tabular}
\caption{ANN AADT Input Features}
\label{table:ANN_AADT_Features}
\end{table}

For training, LA AADT values per vehicle type from the Department for Transport \cite{AADTLocalAuthorityData} are concatenated with count site datasets, serving as ground truth for multi-regression modeling. The vehicle types categories consist of: cars and taxis, LGV's, HGV's, and buses and coaches. LA AADT is computed by selecting the maximum count site vehicle type AADT (by direction) to provide a "worst case" estimation, avoiding systematic under-estimations at the emissions stage. 
 
To ensure unbiased model evaluation, the satellite image acquisition dates to be tested on are all in the 2018 year. Thus, training and validation extend up to 2017, while 2018 is reserved for testing. Each count site has its own model, leveraging diverse traffic flow patterns and ample data (approx. 32,000 data-points per site) for optimal performance. Validation MAPE's at training completion range from 0 to 0.2, with early stopping (patience: 3) to prevent overfitting.

\subsubsection{ANN AADT Prediction}

To make a prediction of LA AADT from a count site, the following steps are done:
\begin{enumerate}
    \item Load the model weights for the LA and count site
    \item Load traffic count prediction and transform using saved min-max parameters
    \item Load historical or estimated speed data (in this evaluation, historical data is used)
    \item Load time data (from satellite image) as month, day and hour
    \item Concatenate transformed traffic counts, speed and time data and feed into model 
\end{enumerate}

Producing an LA AADT for each direction of travel allows for the mean value to be calculated. As some predictions are over-estimates, while others will be under-estimates, this could further improve the accuracy as the mean value will often be closer to the true value. The mean value is used for the metric calculations as it was found to improve results on the majority of count sites chosen, compared to the median or maximum.

\subsection{Motorway LA GHG Emissions} \label{Methods: Road Transport LA GHG Emissions}

This stage uses the LA AADT predictions from the previous stage to estimate motorway LA GHG emissions, representing stage 6 of the pipeline. 
\\ \\
Table \ref{table:ghg_emissions_features} shows the variables and their respective values chosen for calculating GHG emissions. Total length of motorway values are published by the DfT \cite{roadlengthstats}, GHG conversion factors are published by DESNZ and BEIS \cite{UKGovGHGConversionFactors}, fuel consumption is published by DfT \cite{UKGovFuelConsumption} and vehicle licensing statistics are published by the DfT \cite{UKVehicleLicenceStats}.

\begin{table}[H]
  \centering
  \begin{tabular}{@{} lcr @{}}
\toprule
\textit{Total length of motorway} & \textit{km} \\
Luton & 4.18 \\
Blackburn & 12.87 \\
Hounslow & 15.77 \\
Havering & 19 \\
Trafford & 9.98 \\
\hline
\textit{GHG Conversion factors} & \textit{kg CO2e} \\ 
Petrol cars & 2.16 \\
Diesel cars & 2.56 \\
\hline
\textit{Fuel consumption (2012)} & \textit{km/litre} \\ 
Petrol cars & 20 \\
Diesel cars & 23 \\
Petrol LGV's & 18.4 \\
Diesel LGV's & 17.1 \\
HGV's & 3.6 \\
\hline 
\textit{Vehicle Licensing weighting (2017)} & Percentage \\
Petrol & 59 \\
Diesel & 40 \\
\bottomrule
\end{tabular}
\caption{GHG Emissions Features}
\label{table:ghg_emissions_features}
\end{table}

The equation for calculating annual GHG emissions per vehicle type, $i$, in an LA, using vehicle type data is:
\begin{equation}
    \begin{aligned}
        \text{VKT}_i & = \text{AADT}_i * \text{Total Motorway Length} * 365 \\
        \text{Litres}_i & = \text{VKT}_i / \text{Specific Fuel Consumption}_i \\
        \text{GHG Emissions}_i & = \text{Litres}_i * \text{GHG Conversion Factor}
    \end{aligned}
\end{equation}

The total GHG emissions is then calculated as the sum of emissions across all vehicle types. We predict emissions across the same four vehicle categories as for AADT.
 
We use the 2012 fuel consumption values from the DfT \cite{UKGovFuelConsumption}. The relevant fuel types for GHG conversion factors are petrol and diesel, for which we assume the average bio-fuel blend as this is found in normal petrol stations. In addition, the vehicle licensing dataset \cite{UKVehicleLicenceStats} shows that in 2017 petrol cars accounted for 59\% of all licensed cars, while diesel cars accounted for 40\% and electric/hybrid cars were 1\%. This weighting is applied to the conversion factors to represent this fuel type distribution. 

\subsection{Flexible Approaches} \label{Implementation: without Vehicle Type Data}

\subsubsection{Live Speed Estimation} \label{Methods: Live Speed Estimation}

This section describes Stage 3 of the pipeline. If historical live speed data is available, this should be used as it will be the most accurate. However, if live speeds are unknown, we estimate them using principal component analysis (PCA) on the time-lag between MS bands of WorldView-2/3 satellite images. The method, based on Salehi et al. \cite{6145661}, is extended by specifying an area, compactness and rectangular shape criterion to road vehicle detection, rather than using only compactness. Potential objects above a threshold for these criteria are identified as road vehicles. The threshold values are hyper-parameters that are optimized by visual inspection on a subset of randomly chosen count site PC change images.

\subsubsection{Without Vehicle Type Data} \label{Methods: Without Vehicle Type Data}

Predicting LA AADT on a vehicle type basis is used as it was found to improve results compared to using a single AADT value by 10\% on average. However, in relation to the UN SGD goals on Climate Action and Sustainable Cities and Communities, an alternative method is developed to calculate GHG emissions from a single AADT value, for use in regions where data on AADT per vehicle type is not available. The method uses the satellite image to estimate the distribution of vehicle types using object classification, and is as follows:
\begin{enumerate}
    \item Vehicle type mapping: Load vehicle object detection results and convert the vehicle type classes from the xView dataset into valid categories from the UK dataset
    \item AADT assignment: Assign predicted overall AADT proportionally to each detected vehicle type
\end{enumerate}

Table \ref{table:vehicle_type_mapping} shows the specific mappings for each xView vehicle class to the UK class.  The rest of the GHG emissions calculation is the same as with vehicle type data. 

\begin{table}[H]
  \centering
  \begin{tabular}{@{}lc@{}}
    \toprule
    xView Class & UK Dataset Class \\
    \midrule
Passenger vehicle, Small car, \\ Passenger car & Cars and Taxis \\ 
\hline
Pickup Truck, Utility Truck, Truck, \\ Trailer, Truck w/ Box, Trailer, Cargo car & LGV's \\ 
\hline
Cargo Truck, Truck Tractor, \\ Truck w/ Flatbed, Truck w/Liquid & HGV's \\ 
\hline
Bus & Buses and coaches \\
    \bottomrule
  \end{tabular}
  \caption{xView to UK Dataset Class Mapping}
  \label{table:vehicle_type_mapping}
\end{table}
 
This mapping is important as emissions vary significantly by vehicle type, for example HGV's have a fuel consumption of 5.6 times less efficient than petrol consumption, as shown in Table \ref{table:vehicle_type_mapping}. Overall, using AADT vehicle type data is a way of introducing robustness to the noise of using a single satellite image to estimate vehicle type distributions. 

\subsection{Extension to A-Roads and Minor Roads} \label{Implementation: Extension to Other Road Types}

This section represents stage 7 of the pipeline. While the satellite images and 15-minute traffic counts data comes from motorway sections, the methods used for LA AADT and emissions calculations can be extended to other road types.
 
Here, the ANN for LA AADT prediction uses the LA AADT value for A-Roads and Minor Roads respectively as the target variable. Thus, it is trained to learn transformations from motorway traffic data to nearby road traffic (i.e. within the same LA).
 
The following points summarise the required value changes when doing AADT and GHG emissions prediction on other road types. The dataset sources used for motorways are the same as for A-roads and minor roads and can be filtered as appropriate. 
\begin{itemize}
    \item ANN LA AADT prediction target variable and ground truth
    \item Emissions variables (total road length and km/litre )
    \item Emissions ground truth data
\end{itemize}

\section{Results \& Discussion}

As the test set for which we report results in this section, we take archive 2018 WorldView-2/3 satellite images of 5 UK motorway 15-minute traffic count sites, each of which consists in a different LA. The LA's were selected based off passing certain criteria, for example less than 10\% missing values in traffic count data \cite{HighwaysEnglandTrafficFlowData}, satellite image having less than 5\% cloud cover \cite{euspaceimaging}, etc. 
 
Traffic flow distributions can vary significantly for each side of travel, thus for each image we run the pipeline separately for each direction of travel. This produces 2 result values per satellite image and LA, where we write the count site name on the same row as the result (e.g. M1/2557A for Luton). Where we report just one result for a LA, the mean across the 2 results has been taken. 

\subsection{15-Minute Traffic Counts} \label{Results: 15-Minute Traffic Counts}

Table \ref{table:15_minute_traffic_counts_evaluation} shows the RMSE and MAPE error metrics for each LA when using historical speed data. The last row of the table is an average score across the count sites.

\begin{table}[H]
  \centering
  \begin{tabular}{@{} lcr @{}}
\toprule
LA Count Site & RMSE & MAPE \\
\midrule
Blackburn 30361032 & 57 & 0.122 \\
30361033 & 263 & 0.47 \\ 
\hline
Luton M1/2557A & 557 & 0.50 \\ 
M1/2557B & 200.3 & 0.17 \\ 
\hline
Havering M25/5790A & 604 & 0.51 \\ 
M25/5790B & 3.4 & 0.003 \\ 
\hline
Hounslow M4/2188A & 320 & 0.42 \\ 
M4/2188B & 24.5 & 0.031 \\
\hline 
Trafford M60/9083A & 576 & 0.61 \\
M60/9086B & 54.9 & 0.063 \\
\hline 
AVERAGE & 266 & 0.29 \\
\bottomrule
\end{tabular}
\caption{15-Minute Traffic Counts Results}
\label{table:15_minute_traffic_counts_evaluation}
\end{table}

This section will evaluate the methods to produce 15-minute traffic counts from a processed satellite image. 
 
Overall, the pipeline to produce 15-minute traffic counts from a single satellite image builds upon previous research in this area. However, the method of PCA-based live speed estimation is novel, and represents an advancement towards the project goals where speed data may not be available.

\subsection{Road Transport LA AADT} \label{Results: Road Transport LA AADT}

Table \ref{table:la_aadt_evaluation} shows the RMSE and MAPE evaluation metrics for motorway and A-road LA AADT results using true speed data.

\begin{table}[H]
  \centering
  \begin{tabular}{@{} lccc @{}} 
\toprule
LA & Road Type & RMSE & MAPE \\
\midrule
Blackburn & Motorways & 4433 & 0.13 \\
& A-Roads & 3533 & 0.20 \\ 
\hline
Havering & Motorways & 5779 & 0.087 \\ 
& A-Roads & 5442 & 0.11 \\
\hline
Luton & Motorways & 703 & 0.0099 \\ 
& A-Roads & 2131 & 0.08 \\ 
\hline
Hounslow & Motorways & 7714 & 0.15 \\
& A-Roads 4253 & 0.07 \\
\hline
Trafford & Motorways & 758 & 0.013 \\
& A-Roads 1710 & 0.05 \\
\hline
AVERAGE & Motorways & 3878 & 0.077 \\
& A-Roads & 3414.0 & 0.10 \\
\bottomrule
\end{tabular}
\caption{Motorway LA AADT Results}
\label{table:la_aadt_evaluation}
\end{table}

A-Roads show traffic data distributions similar to motorways, resulting in comparable results, with A-Roads having a MAPE only ~0.025 worse than motorways. Comparing to Ganji et al.'s study on AADT prediction in Toronto, Canada, achieving an R$^2$ value of 0.78, our base pipeline focuses on motorways and achieves an R$^2$ of 0.92. The ANN model handles 15-minute traffic counts effectively. No similar UK studies are available.
 
\begin{figure}[H]
    \centering
    \includegraphics[width=0.7\textwidth]{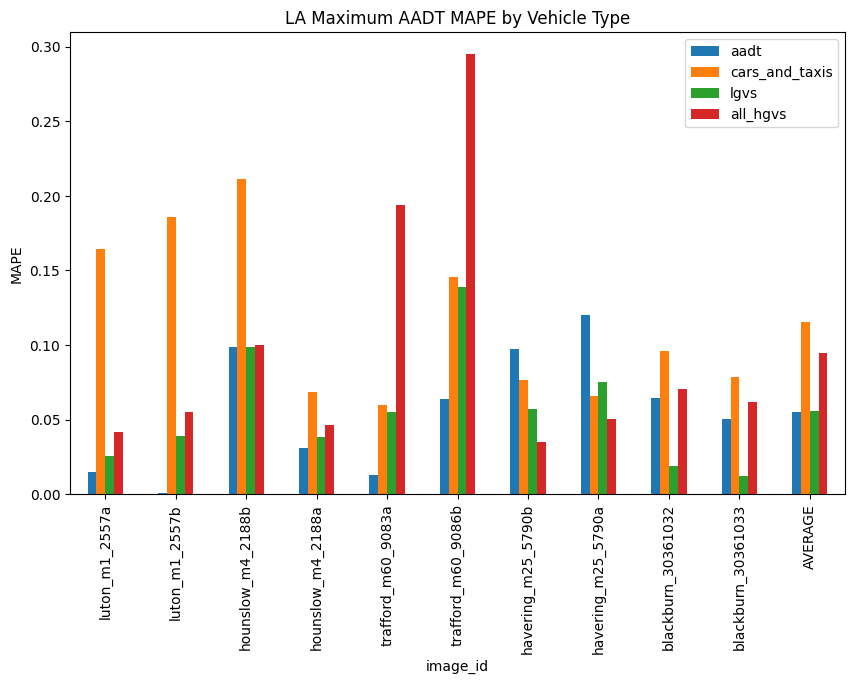}
    \caption{AADT MAPE by Vehicle Type Comparison}
    \label{fig:aadt_mape_by_vehicle_type_comparison}
\end{figure}

Figure \ref{fig:aadt_mape_by_vehicle_type_comparison} displays AADT MAPE values by vehicle type. Note that the buses and coaches class is excluded due to significantly worse performance. Their average MAPE is 0.8, likely due to data imbalance. LGVs have similar MAPE values to overall AADT, while cars and taxis and HGVs have approximately double MAPE values, suggesting LGV traffic patterns on motorways are less noisy. The pipeline's limitation is the omission of surrounding area variables, which could impact cars and taxis' MAPE.

\subsection{Road Transport LA GHG Emissions} \label{Results: Road Transport LA GHG Emissions}

Table \ref{table:la_ghg_evaluation} shows the Motorway and A-Road RMSE and MAPE evaluation metrics for annual LA GHG Emissions prediction, where we report an average MAPE of 0.24 for motorways and 0.19 for A-Roads. Interestingly, A-Roads is slightly lower than motorway's. The distribution of results is more variable than with AADT, with the smallest being 0.08 and the highest being 0.33.  

\begin{table}[H]
  \centering
  \begin{tabular}{@{} lccc @{}} 
\toprule
LA & Road Type & RMSE & MAPE \\
\midrule
Blackburn with Darwen & Motorways & 8.32 & 0.21 \\
& A-Roads & 14.7 & 0.27 \\ 
\hline
Havering & Motorways & 32.8 & 0.20 \\
& A-Roads & 32.8 & 0.08 \\ 
\hline
Luton & Motorways & 6.61 & 0.19 \\ 
& A-Roads & 18.0 & 0.33 \\ 
\hline
Hounslow & Motorways & 6.66 & 0.10 \\
& A-Roads & 22.6 & 0.12 \\
\hline
Trafford & Motorways & 45.7 & 0.47 \\
& A-Roads & 14.9 & 0.16 \\
\hline
AVERAGE & Motorways & 20.0 & 0.24 \\
& A-Roads & 16.5 & 0.19 \\
\bottomrule
\end{tabular}
\caption{Motorway LA GHG Emissions Results}
\label{table:la_ghg_evaluation}
\end{table}

\begin{figure}[H]
    \centering
    \includegraphics[width=0.6\textwidth]{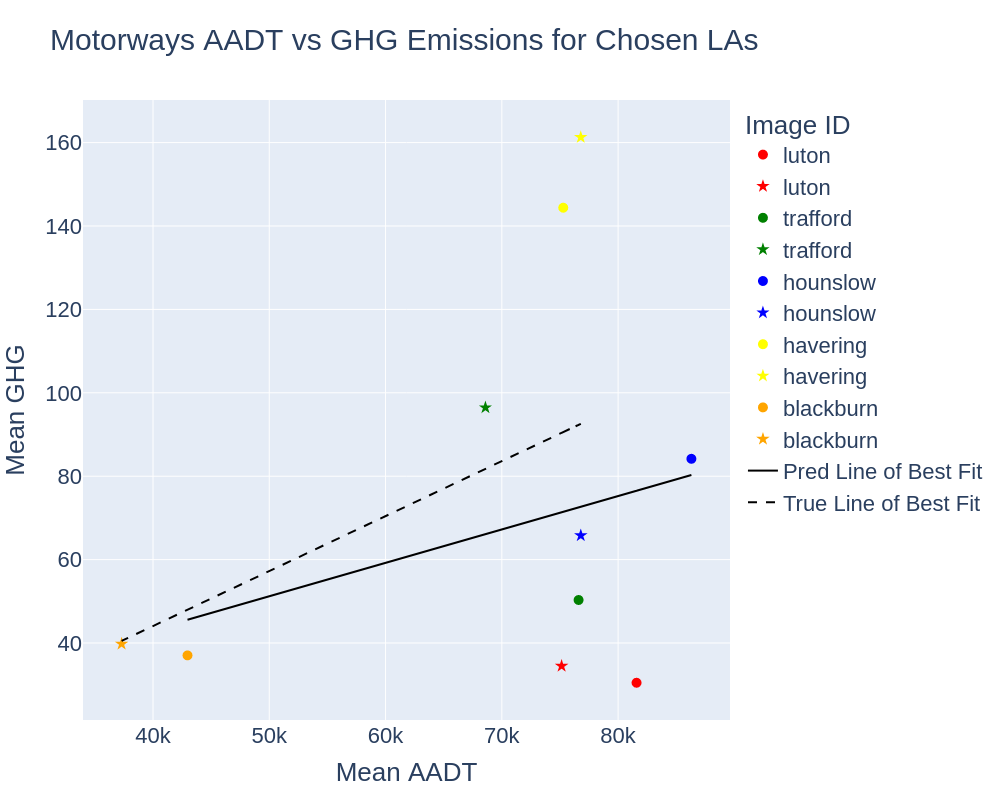}
    \caption{Motorway AADT and GHG Emissions Scatter Plot}
    \label{fig:motorways_aadt_ghg_eval_max_plot}
\end{figure}

\begin{figure}[H]
    \centering
    \includegraphics[width=0.6\textwidth]{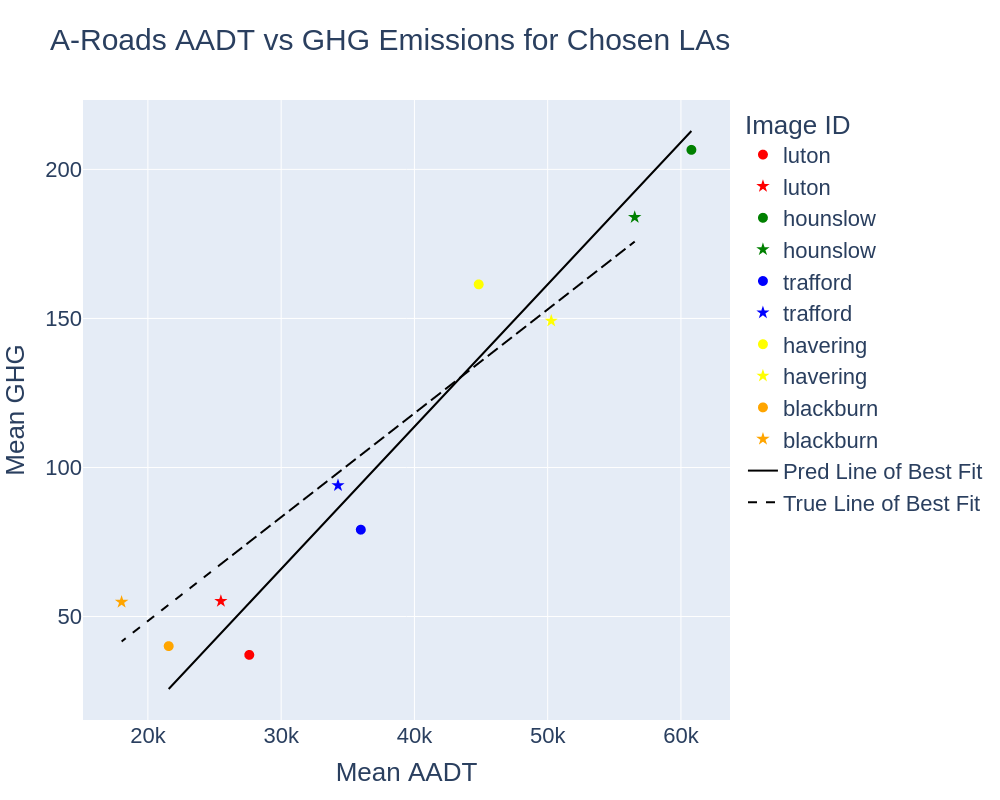}
    \caption{A-Roads AADT and GHG Emissions Scatter Plot}
    \label{fig:aroads_aadt_ghg_eval_max_plot}
\end{figure}

For similar reasons to AADT, direct comparisons to other state-of-the-art emissions predictions is difficult. Ganji et al. \cite{ganji2020methodology} is perhaps the closest study, where they predict emissions in 2015 for the city of Toronto, Canada to be 6.0823 million tonnes/yr, while the Atmospheric Fund (TAF) for 2015 for Toronto was 5.25 million tonns/yr (TAF, 2018) \cite{ganji2020methodology}. This represents an MAPE of 0.15. There are no public studies similar to ours conducted in the UK. 
Figure's \ref{fig:motorways_aadt_ghg_eval_max_plot} and \ref{fig:aroads_aadt_ghg_eval_max_plot} display the predicted and true LA AADT and GHG emissions Motorways and A-Roads respectively. For predictions, we display the mean value across the two count sites in each LA. Circle markers represent predictions, while star markers represent true values. The results show that the pipeline is able to accurately estimate LA GHG road emissions on motorways using AADT values per vehicle type. It is worth noting that A-roads exhibit a stronger correlation between AADT and GHG emissions than motorways.

\subsection{Flexible Approaches}

\subsubsection{Live Speed Estimation}

\begin{figure}[H]
    \centering
    \includegraphics[width=0.3\textwidth, height=5cm]{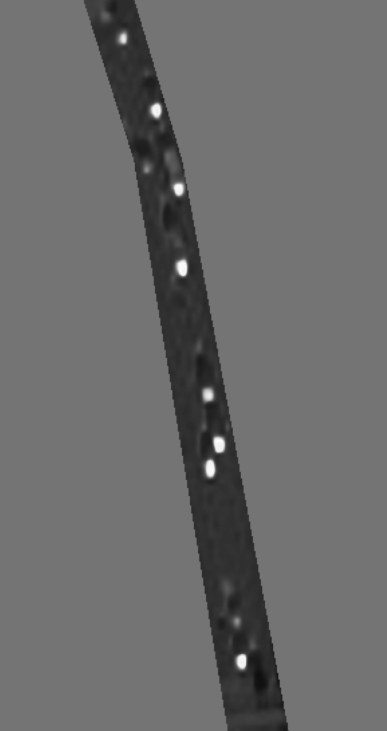}
    \includegraphics[width=0.3\textwidth, height=5cm]{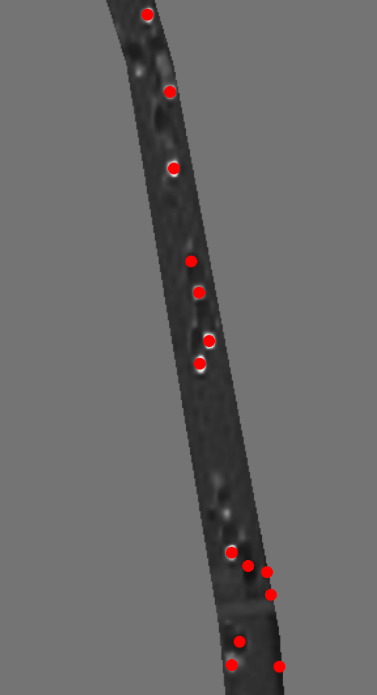}
    \caption{Zoomed Section of Havering PC Change Image (left) and Vehicle Centroids Detection (Right)}
    \label{fig:havering_pc_change_img_and_vehicle_centroids}
\end{figure}

Figure \ref{fig:havering_pc_change_img_and_vehicle_centroids} shows a zoomed in section of the PC change image for Havering M25/5790A (left) as well as the detected vehicle centroids, indicated by red dots (right).

Table \ref{table:live_speed_estimation_evaluation} shows the MAPE of live speed estimation against historical speeds, as well as 15-minute traffic count MAPE results when using live speed estimation instead of true speeds. Where a count site shows an MAPE of 1.0 or N/A, this means no vehicles were paired, i.e. a bright and dark spot were not found within a threshold value of each other. This threshold value corresponds to the maximum distance a car could travel at 70mph in the time lag. Visual inspection of the images show that they all contain vehicles and so at least one vehicle pair should be found. Thus, this represents a failed stage in the live speed estimation method. 

\begin{table}[H]
\centering
\begin{tabular}{@{} lccc @{}} 
\toprule
LA Count Site & Speed MAPE & Traffic Count MAPE \\
\midrule
Blackburn 30361032 & 0.21 & 0.04 \\ 
30361033 & 1.0 & N/A \\
\hline
Luton M1/2557A & 1.0 & N/A \\
M1/2557B & 0.02 & 0.21 \\
\hline
Havering M25/5790A & 0.05 & 0.48\\ 
M25/5790B & 0.39 & \\
\hline
Hounslow M4/2188A & 1.0 & N/A \\
M4/2188B & 0.01 & 0.03 \\
\hline
Trafford M60/9083A & 0.3 & 0.49 \\
M60/9086B & 0.05 & 0.11 \\
\hline
AVERAGE & 0.3 & 0.48 \\
\bottomrule
\end{tabular}
\caption{Live Speed Estimation Results}
\label{table:live_speed_estimation_evaluation}
\end{table}

Live speed estimation can produce MAPE values of less than 0.05, for example the Luton count site M1/2557A. However, it can also fail to produce an estimate for some images. A possible reason and disadvantage for PCA-based speed estimation is the level of hyper-parameter tuning. This will be a significant consideration as the pipeline is scaled in order to meet the project goals. In the original paper \cite{6145661}, only one image was used, thus parameter tuning was simpler. 

\subsubsection{Without Vehicle Type Data}

\begin{figure}[H]
    \centering
    \includegraphics[width=0.7\textwidth]{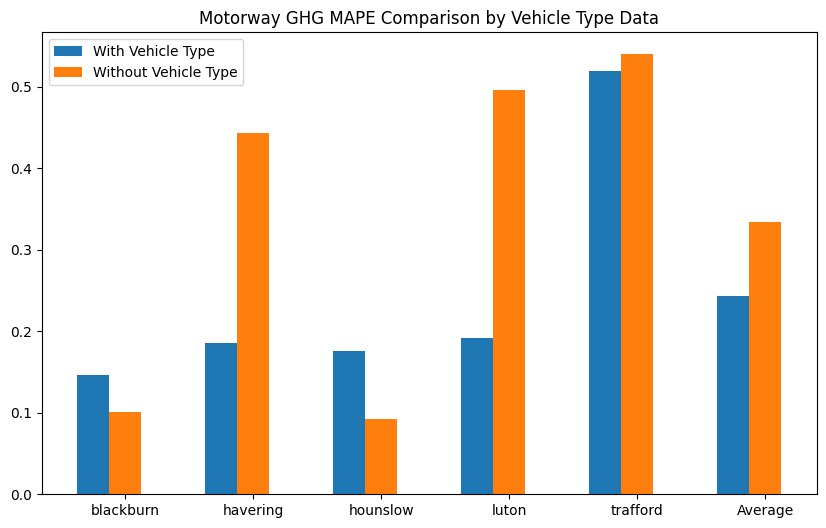}
    \caption{MAPE Comparison for emissions calculations without Vehicle Type Data}
    \label{fig:motorways_aadt_ghg_eval_no_vehicle_type_plot}
    
\end{figure}

Figure \ref{fig:motorways_aadt_ghg_eval_no_vehicle_type_plot} shows the bar plot comparison of AADT and GHG emissions with and without vehicle type data, for motorways. This uses the method as described in Section \ref{Methods: Without Vehicle Type Data} instead of the one in Section \ref{Methods: Road Transport LA GHG Emissions}.

As expected, not using AADT on a vehicle type basis reduces the accuracy in the emissions calculations, with the average MAPE reducing by around 0.1. Trafford remains as an outlier in its' MAPE, while Luton and Havering both see increases of 0.2 and 0.3 MAPE respectively for without vehicle type data. 

\section{Conclusion}

In conclusion, we introduce a novel two-part pipeline. It starts with a raw satellite image containing a motorway section. The first part predicts traffic counts at count sites within the image, as depicted in Figure \ref{fig:full_pipeline_diagram}. The second part estimates Average Annual Daily Traffic (AADT) and Greenhouse Gas (GHG) emissions at the Local Authority (LA) level.
 
This work provides several novel contributions. It is the first to predict AADT and GHG emissions at a regional level from satellite imagery in the UK. In addition, it takes full advantage of the extensive public and free UK datasets for predicting these emissions by categorising AADT into four vehicle type categories. In addition, we estimate across the three major road types: Motorways, A-Roads and Minor Roads, using estimated 15-minute motorway traffic counts. This allows for novel insights and comparisons into traffic and emissions patterns. Finally, we extend Salehi et al. \cite{6145661} and introduce a method for estimating AADT by vehicle type to provide additional data flexibility to the pipeline. 
 
The goal of this work is to support the progress toward UN Sustainable Development Goals, notably Climate Action and Sustainable Cities and Communities.

\section{Acknowledgments}

We would like to thank the European Space Agency (ESA) for kindly providing the satellite data used in this paper. 

Dr. Baiz would like to thank Prof. Julie McCann for hosting Dr. Baiz in the Adaptive Emergent Systems Engineering (AESE) group at Imperial College London.

\bibliographystyle{unsrt}  
\bibliography{references}

\end{document}